\documentclass[12pt]{article}
\usepackage[margin=1in]{geometry}
\usepackage{graphicx}
\usepackage{subcaption}
\usepackage{amsmath,amssymb}
\usepackage{authblk}
\usepackage{cite}
\usepackage{lineno}
\usepackage[colorlinks=true, linkcolor=blue, citecolor=blue, urlcolor=blue]{hyperref}

\title{Temperature-Tunable Entangled Photon Source for Multiplexed Time-Resolved Fluorescence on a Nanophotonic Platform}

\author[1]{Ayantika Sengupta}
\author[2]{Nathan A. Harper}
\author[1]{Emily Y. Hwang}
\author[2]{Scott K. Cushing\thanks{scushing@caltech.edu}}

\affil[1]{Department of Applied Physics and Materials Science, California Institute of Technology, Pasadena, California 91125, USA}
\affil[2]{Department of Chemistry and Chemical Engineering, California Institute of Technology, Pasadena, California 91125, USA}

\date{}

\begin{document}
\maketitle

\begin{abstract}
Compact, scalable, and multiplexed fluorescence lifetime sensors are of great interest for point-of-care diagnostics. However, current solutions either lack broad-range wavelength-tuning capabilities or involve complex optical setups that hinder miniaturization. On-chip entangled photon sources offer a promising alternative for time-resolved spectroscopy with their strong temporal correlations, tunable spectral characteristics, and small footprints. Here, we develop a temperature-tunable, visible quantum light source on thin-film lithium niobate (TFLN) with a continuous tuning range greater than one octave, spanning 564.5~nm to 1.494~$\mu$m using only one waveguide. The tunability is enabled by utilizing type-I phase matching. We measured an on-chip efficiency of \boldmath$(3.88\pm0.20)\times10^{9}$ pairs/s/mW, comparable to the most efficient type-0 bulk lithium niobate sources. These results show that the TFLN platform is ideal for on-chip integrated photonic and multiplexed lifetime imaging and sensing.
\end{abstract}

Compact, cost-effective, and high specificity optical biosensors capable of detecting a diverse range of biomarkers could be instrumental for point-of-care diagnostics. Time-resolved fluorescence measurements offer the ability to distinguish spectrally overlapping fluorophores\cite{datta2020fluorescence} and overcome limitations of traditional intensity-based methods, such as signal fluctuations arising from changes in illumination intensity or fluorophore concentration\cite{becker2012fluorescence}. Fluorescence lifetime sensing is widely used to measure cellular microenvironments\cite{sanders1995quantitative}, including ion concentrations, pH and oxygenation, to monitor protein interactions and to advance cancer diagnostics and drug-delivery studies\cite{karrobi2023fluorescence}. Integrating fluorescence lifetime measurements with excitation-wavelength multiplexing can probe multiple biological targets, expanding biosensing capabilities for clinical applications. However, designing fluorescence lifetime spectrometers that are compact and widely tunable across the visible and near-infrared (NIR) regions remains challenging. To date, excitation-wavelength multiplexed fluorescence lifetime sensing has relied on complex optical setups, including synchronous switching between multiple excitation channels \cite{suarez2021metabolic}, Fourier transform frequency-domain methods using Michelson interferometers\cite{zhao2012quantitative}, interferometric excitation combining two broadband pulses with variable time delay\cite{maly2024interferometric}, multiple nonlinear mixing optics\cite{periasamy1996time}, or supercontinuum lasers with spectral filters \cite{volz2018white}. Although compact fluorescence lifetime setups have benefited from the development of on-chip pulsed lasers and pulsed LEDs, these sources lack wide-range excitation wavelength tunability\cite{kennedy2008fluorescence,8630989}. Recently, entangled photons produced through spontaneous parametric downconversion (SPDC) were used to determine fluorescence lifetimes using the temporal correlations present in the photon pairs \cite{harper2023entangled,eshun2023fluorescence}. Entangled photons provide a simple yet effective method of wavelength tuning by changing the temperature of the SPDC source. In addition to offering a much smaller footprint than conventional sources for lifetime spectrometers, SPDC source spectra can be engineered to be narrowband or broadband\cite{xin2022spectrally}. Narrowband sources can increase the maximum number of multiplexing channels possible. While entangled photon sources are less bright than their classical counterparts, their flux is high enough to easily saturate single-photon counting detectors and cameras. This has led to a rapidly growing interest in exploring entanglement for spectroscopy and imaging techniques\cite{roslyak2009nonlinear}, such as stimulated Raman spectroscopy\cite{dorfman2014stimulated}, sub-diffraction resolution microscopy\cite{defienne2022pixel}, quantum optical coherence tomography\cite{graciano2019interference}, and phase contrast microscopy\cite{ono2013entanglement}. Developing miniaturized, tunable SPDC sources could advance the cutting edge of quantum light development for chemical or biological imaging and sensing approaches.

\begin{figure}[htbp]
\includegraphics[width=12cm]{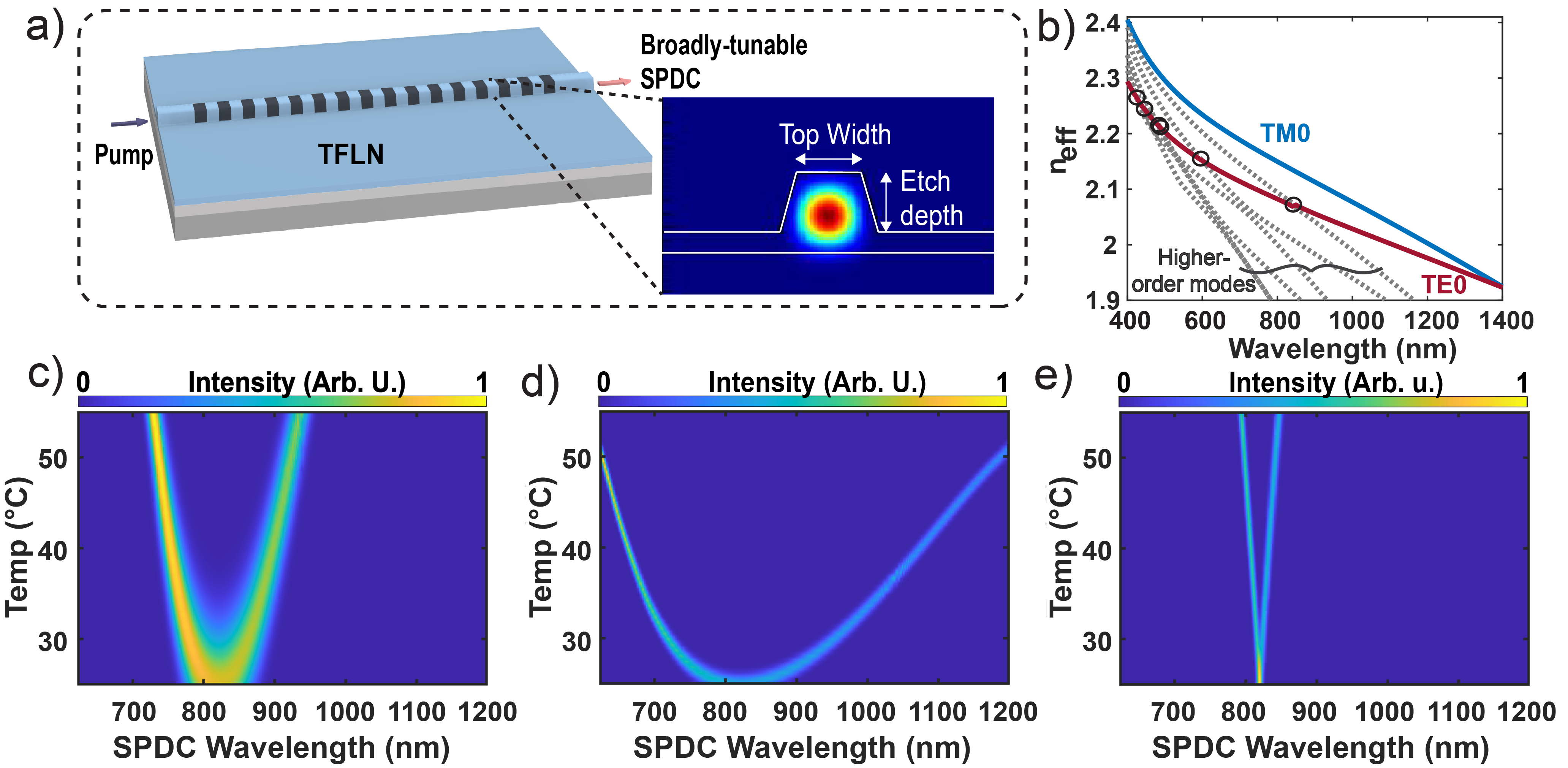}
\centering
\caption{(a) Schematic of periodically poled lithium niobate nanophotonic waveguide for SPDC generation. Inset shows the simulated rib waveguide geometry and mode profile for comparison of phase matching types. (b) Dispersion curves showing mode crossings, indicated with black circles, of the quasi-TE0 mode with higher order modes. Temperature tuning curves for (c) type-0 (d) type-I (e) type-II phase matching.}
\label{fig1}
\end{figure}

Thin-film lithium niobate (TFLN) is a promising candidate for time-resolved fluorescence due to its broad transparency window down to $\sim$~330~nm, high second-order nonlinearity, low optical losses and scalability\cite{boes2023lithium}. The strong nonlinear interaction in TFLN from sub-$\mu$m modal confinement and high-index contrast has led to significant advancements in highly efficient SPDC sources \cite{harper2024highly,fang2024efficient,xue2021ultrabright,javid2021ultrabroadband,fang2025high}. These sources achieve pair generation rates that are orders of magnitude higher than traditional nonlinear crystals like BBO\cite{zhang2021spontaneous,sansa2022visible}, as well as bulk lithium niobate crystals\cite{kuo2020demonstration,prabhakar2020two,szlachetka2023ultrabright} and micron-scale diffused and ridge waveguides\cite{fujii2007bright,bock2016highly,jechow2008high,zhang2023scalable}. SPDC studies on TFLN are more common in the telecom regime; however, recent advancements have shown that sources in the visible-NIR can be achieved with brightness on the order of $10^9$ pairs/s/mW/nm\cite{harper2024highly}, enabling biosensing applications with fluorescent dyes. To further optimize these sources for fluorescence lifetime measurements, the tunability of the source must be increased to reach the visible part of the spectrum.

In this work, we develop an integrated photonic source of visible entangled photon pairs using thin-film lithium niobate. We achieve a continuous tuning range greater than one octave, spanning 564.5 nm to 1.494 $\mu$m using only one waveguide. Compared to most TFLN devices to date, which use type-0 phase matching for high efficiency downconversion, we find that type-1 phase matching provides $\sim$~4.7 times better temperature tunability than type-0 due to the thermal birefringence of lithium niobate. Furthermore, mode crossings, which usually disrupt wide tuning in type-0 devices, are eliminated with type-I phase matching by producing pairs in the fundamental quasi-TM mode. Due to the tight modal confinement of the thin-film platform, a high efficiency of $(3.88\pm0.20)\times10^{9}$ pairs/s/mW is achieved, comparable to the most efficient bulk type-0 sources despite the lower nonlinear coefficient of type-I downconversion. With a narrow bandwidth of $<15$~nm far from degeneracy, this source can distinguish between different biomarkers and can be further improved by compensating for thickness variations in the thin-film lithium niobate layer.

The spectral and tuning characteristics of an SPDC source are highly dependent on the dispersion of the nonlinear crystal, as this affects which wavelengths will be phase matched\cite{karan2020phase}. Depending on the polarization of the interacting pump, signal, and idler fields, three types of phase matching can be satisfied, namely, type-0, type-I, and type-II. The following polarizations are specific to X-cut TFLN waveguides, and the quasi-TE and quasi-TM modes are represented as TE and TM. In type-0, the pump, signal, and idler photons share the same TE polarization. In type-I, the TE pump is orthogonally polarized to the TM downconverted photons. In type-II, the TE signal and TM idler are orthogonally polarized to one another. The dispersion of the X-cut rib waveguide geometry in Fig.~\ref{fig1}a was simulated using a finite-difference Maxwell equation solver software to compare the three types of downconversion. As can be seen in Fig.~\ref{fig1}b, the fundamental TM mode of the X-cut waveguide exhibits no mode crossings across a broad range in the visible and NIR because the ordinary refractive index ($n_o$) is greater than the extraordinary ($n_e$) for lithium niobate. Meanwhile, the fundamental TE mode crosses with several higher-order TM modes\cite{pan2019fundamental}, which can only be eliminated by reducing the dimensions of the waveguide close to the wavelength of light to enhance waveguide dispersion. 

The temperature-dependent spectral response was modeled using the Sellmeier coefficients for bulk lithium niobate\cite{gayer2008temperature} in the mode solver. The temperature tuning spectra of the three types of SPDC are shown in Figs. 1c-e. In lithium niobate, $n_e$ changes at a higher rate with temperature compared to $n_o$. Thus, for an extraordinary pump (TE) and ordinary signal and idler (TM) modes, as in the case of type-I, the thermo-optic birefringence has the most pronounced effect, making it especially sensitive to temperature variations. For a quantitative comparison, the tuning of the second harmonic generation wavelength with temperature was calculated as 49 pm/°C, 230 pm/°C and 40 pm/°C for type-0, type-I and type-II, respectively. These values suggest approximately 4.7 times greater temperature tunability in type-1 phase matching than possible with type-0.

\begin{figure}[htbp]
\includegraphics[width=8cm]{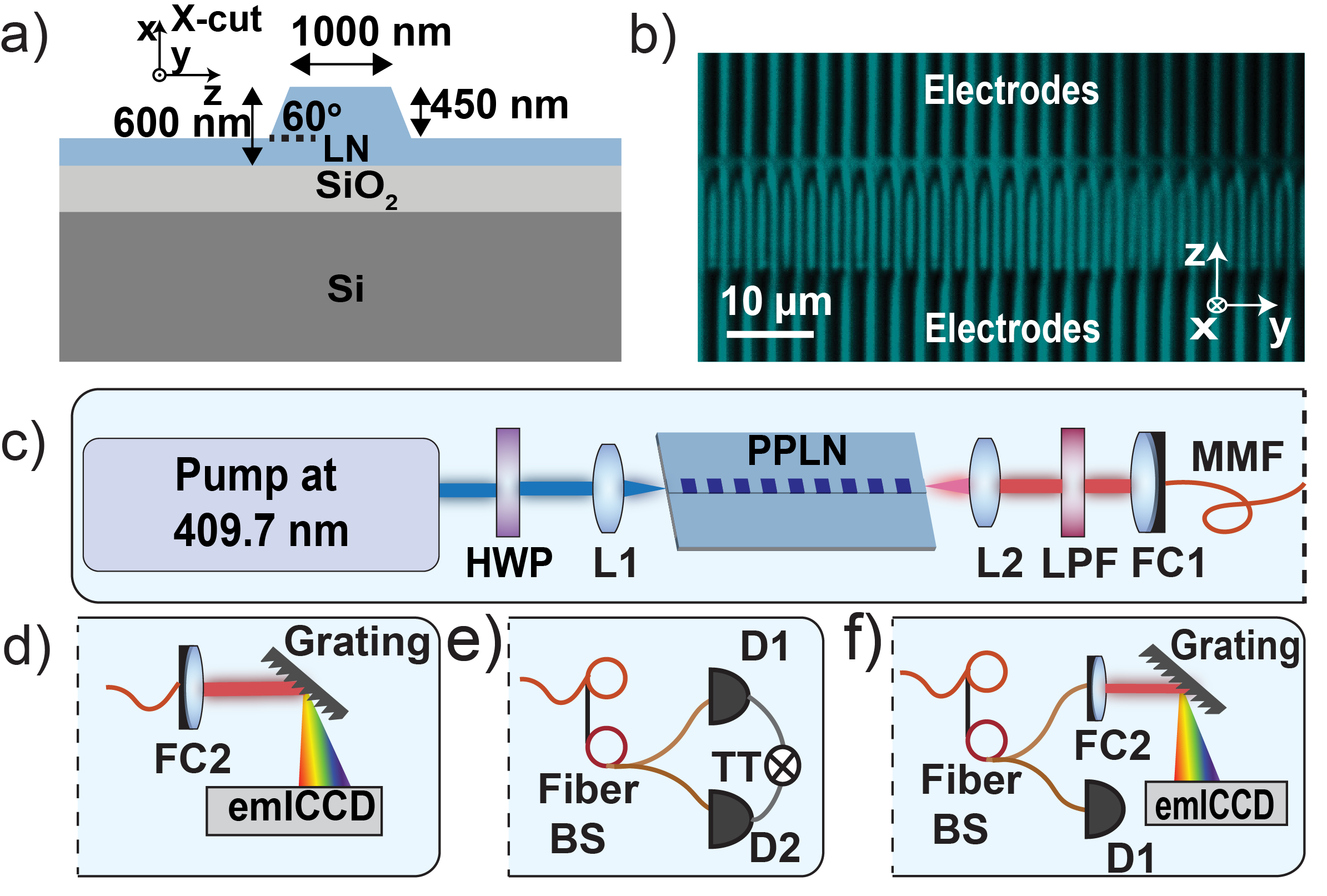}
\centering
\caption{(a) Waveguide dimensions. (b) Second-harmonic microscopy image of the poled region. Optical setups for (c) pump coupling and SPDC collection, (d) characterizing photon pair spectra, (e) coincidence counting and, (f) characterizing temperature-tuned spectra. HWP, half-wave plate; L1, aspheric lens; L2, reflective objective; LPF, long-pass filter; FC1 \& FC2, fiber couplers; MMF, multimode fiber; Fiber BS, fiber beam spliter, D1 \& D2, SPADs; TT, time-tagger.}
\label{fig2}
\end{figure}

The geometric parameters of the waveguide are shown in Fig.~\ref{fig2}a. A top width of 1000~nm, an etch depth of 450~nm, and a total lithium niobate thickness of 600~nm are used. The geometry was selected to avoid mode crossings near the 409.7 nm pump wavelength. At room temperature, a poling period of 2.8 $\mu$m is predicted to give first-order quasi phase matching for type-I SPDC. We fabricate waveguides with a 7~mm poled length from a 1~$\times$~1~cm X-cut TFLN die (NanoLN). Detailed fabrication procedures can be found in Ref.\cite{harper2024highly}. The resulting poled structure was imaged using second harmonic microscopy, as shown in Fig.~\ref{fig2}b, and suggests a duty cycle of approximately 41\%. Atomic force microscopy confirmed that the fabricated design closely matches the intended geometry and the sidewall angle is $60 ^\circ$, resulting from argon plasma etching.

\begin{figure}[!t]
\centering
\includegraphics[width=10cm]{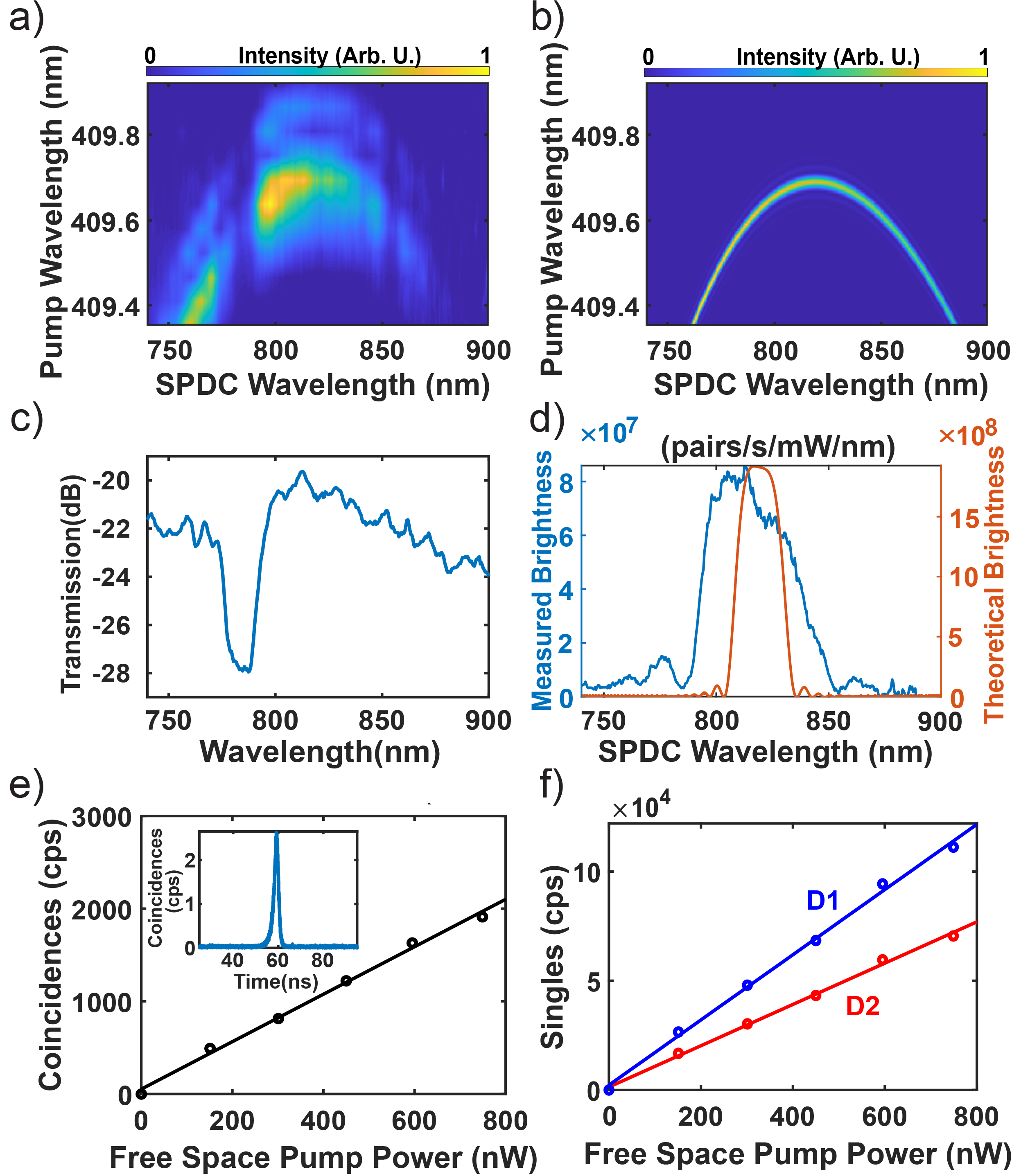}
\caption{(a) Measured and (b) theoretical entangled photon spectra as a function of pump wavelength. (c) Transmission spectrum measured through the waveguide of interest. (d) Measured and theoretical brightness. The measured brightness is determined by scaling the degenerate SPDC spectrum lineout so that its integration over wavelengths equals the observed efficiency. Rate of (e) coincidences after subtracting accidentals and (f) singles as a function of pump power in counts per second (cps) for detectors D1 (blue) and D2 (red). Inset in Fig.~\ref{fig3}e shows a coincidence histogram recorded at a free space pump power of 750~nW.}
\label{fig3}
\end{figure}

The SPDC spectrum is initially characterized at room temperature. The experimental setups are depicted in Figs. 2c-d. The waveguide is pumped using a tunable, single-frequency continuous-wave Ti:sapphire laser (Matisse C) that has been frequency-doubled with a BBO crystal(Eksma Optics BBO-605H). The laser is coupled to the waveguide input facet using an aspheric lens (NA = 0.58, Thorlabs C140TMD-A), and the generated SPDC is collected off-chip using a reflective objective (Thorlabs LMM40XF-UVV) to avoid chromatic aberrations. The SPDC is spectrally filtered to eliminate the pump wavelength, coupled into a multimode fiber (Thorlabs M122L01), and directed to a grating spectrometer where it is detected by an electron multiplying intensified CCD camera (emICCD). The pump wavelength is swept while recording the SPDC spectrum to investigate the phase matching quality. Fig.~\ref{fig3}a shows the recorded entangled spectra at each input wavelength normalized by the pump power and grating efficiency. The theoretical emission as a function of pump wavelength is shown in Fig.~\ref{fig3}b. The spectral broadening and additional weaker peaks in Fig.~\ref{fig3}a are not predicted by theory and are likely due to thickness variations in the lithium niobate layer along the waveguide length\cite{xue2021effect}. The lower than expected intensity observed at the idler wavelengths may be explained by the sharp decline in camera sensitivity above 870 nm. The dip in intensity at around 785 nm and its conjugate wavelength at 864 nm is probably due to absorption losses from impurities in lithium niobate. This is further investigated by recording the transmission spectrum of an unpolarized white light source through the chip, which shows a clear dip around 785~nm, as depicted in Fig.~\ref{fig3}c.

The coincidences between two single photon avalanche diodes (SPADs) are measured to characterize the device efficiency. The SPDC output is divided by a 50:50 fiber beam splitter (Thorlabs TM105R5F1A), with each fiber connected to a SPAD (Laser Components Count), as shown in Fig.~\ref{fig2}e. A time tagger (Picoquant PicoHarp 300) records coincidences and singles rates while the pump power is increased using a variable neutral density wheel. The resultant plots are shown in Figs. 3e and 3f. Coincidences are measured over an 8 ns time window centered on the peak. Accidentals are calculated from a separate 8 ns window away from the peak and subtracted from the coincidences. The singles counts are recorded for 100 ms and averaged over 100 measurements. A coincidence histogram recorded with 0.75 $\mu W$ free space pump power is depicted in the inset of Fig.~\ref{fig3}e. The entangled pair generation rate is calculated from the linear fits of the singles and coincidence rates as a function of pump power and corrected for detector efficiencies using the formalism described in Ref.\cite{harper2024highly}. In these calculations, it is assumed that 70.9\% of the free space pump power couples into the waveguide mode, as predicted by the mode solver, resulting in an on-chip efficiency of $(3.88\pm0.20)\times10^{9}$ pairs/s/mW. However, this is a lower bound of the efficiency, as the actual on-chip pump power is expected to be lower than theoretical predictions due to imperfect pump coupling and propagation losses. We calculate the actual pump coupling efficiency to be closer to 7.3 dB, estimated from measured transmission losses. This results in an efficiency of $(1.48\pm0.08)\times10^{10}$ pairs/s/mW. A comparison of the measured and theoretical brightness of the degenerate spectrum is shown in Fig.~\ref{fig3}d. The FWHM bandwidth is approximately 43~nm, compared to a theoretically predicted bandwidth of 17.2~nm. The spectral broadening due to thickness variations results in a reduced peak brightness of $(8.59\pm0.44) \times 10^7$~pairs/s/mW/nm. Furthermore, photon pair losses impact the heralding efficiency, excitation rate and integration times required to measure accurate lifetimes. Therefore, loss characterization is essential to evaluate the sensitivity and detection limits of SPDC-based biosensors. In our setup, the losses incurred in the two paths from the waveguide to the detectors are 9.8 dB and 11.8 dB, as deduced from the coincidence to singles ratio. These values account for losses from the chip as well as from all the downstream optics. Separate measurements of the fiber coupling efficiency and detector efficiency suggest a chip-to-free-space output coupling loss of 8.2 dB. The transmission losses measured from the chip are 14.6 dB and 14.2 dB at the pump and SPDC degenerate wavelengths, respectively.

\begin{figure}[htbp]
\centering
\includegraphics[width=8cm]{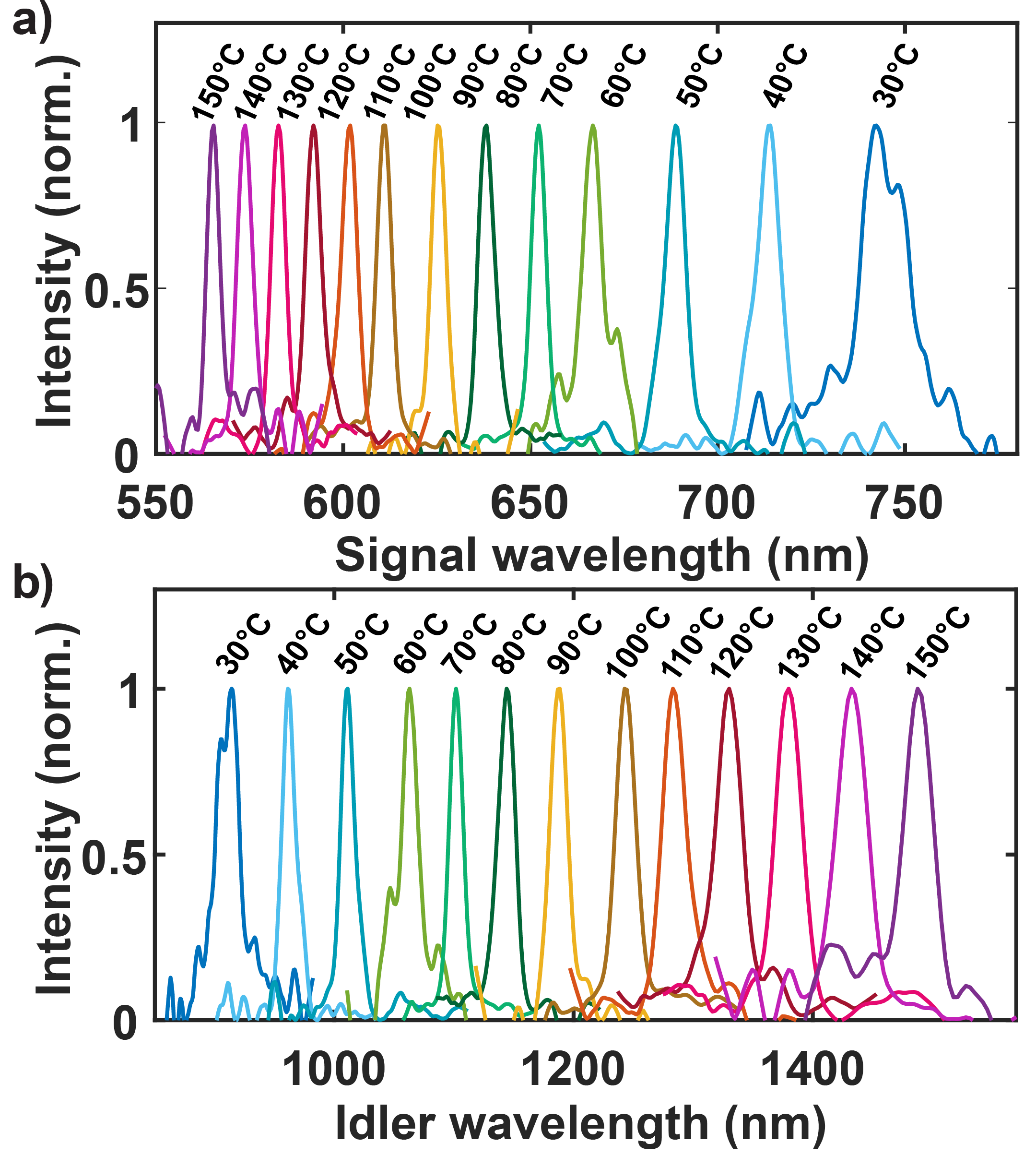}
\caption{Lineouts of the (a) measured signal spectra and (b) simulated corresponding idler spectra shown at 10 $^\circ$C temperature increments. Note that all the nondegenerate peaks have FWHM bandwidths below 15 nm. }
\label{fig4}
\end{figure}

Finally, the wavelength tunability is verified by varying the temperature of the waveguide. The TFLN chip is mounted on a temperature-controlled crystal oven with the optical setup shown in Fig.~\ref{fig2}f. At each temperature, the device coupling is optimized using a single SPAD connected to one port of the 50:50 beamsplitter. The other arm of the beamsplitter is spectrally resolved with a grating spectrometer and imaged in an emICCD. The pump wavelength and off-chip power are constant at 409.7 nm and 0.95 $\mu$W, respectively. At room temperature, degenerate SPDC photons are generated. Subsequently, the temperature is increased in steps of 10$^\circ$C from $30^\circ$C to $150^\circ$C. Fig.~\ref{fig4}a shows the broadly tuned SPDC emission as a function of temperature. Each spectrum has been normalized to unity. As expected from the temperature-tuned spectra in Fig.~\ref{fig1}d, the wavelength shift per degree becomes smaller at shorter wavelengths due to the steepening of the tuning curve. Although spectra at wavelengths longer than 950~nm could not be collected due to the camera sensitivity in the IR, the conjugate spectra of the generated idler photons (Fig.~\ref{fig4}b) are estimated based on energy conservation from the measured signal spectra.  The idler extends as far as 1494~nm for the spectrum at 150$^\circ$C . Therefore, the source achieves a total spectral coverage of 929.5 nm, from 564.5 nm to 1494 nm. 

The tunability of this source offers great promise in simplifying instrumentation for multiplexed biosensing since a single waveguide can generate emission spanning wider than an octave in the visible and NIR range. Moreover, the FWHM of the nondegenerate SPDC spectra remains under 15~nm at all temperatures, which is considerably narrower than the typical absorption profile of dyes in solution. Thus, operating the waveguide at nondegeneracy guarantees narrowband emission, ideal for specific targeting of fluorophores without needing additional spectral filtering. While the source bandwidth is broader than that of laser diodes or spectrally filtered supercontinuum sources, it is narrower than commercial LEDs. In future designs, adapted poling\cite{chen2024adapted} can be implemented to compensate for waveguide thickness variations to approach the theoretical limit of $<5$~nm, placing the source specificity on par with conventional excitation sources. The combination of sufficiently high photon flux to saturate single-photon detectors, broad spectral tunability, narrow emission bandwidths, and strong temporal correlations opens opportunities for selective, multiplexed, and low-background fluorescence measurements in biological samples. This work lays the foundation for distributable continuous wave-pumped quantum-enhanced sensing, particularly in time-resolved techniques, which conventionally require pulsed or modulated lasers.

\bibliographystyle{unsrt}
\bibliography{references}
\end{document}